\documentclass[11pt,english,a4paper]{revtex4}
\usepackage[T1]{fontenc}
\usepackage[latin1]{inputenc}

\makeatletter

\usepackage{latexsym}

\usepackage{babel}
\makeatother
\begin{document}

\title{\textmd{\Large Thermodynamic Interpretation of the Field Equations
of BTZ Charged Black Hole near the Horizon}}

\author{Eduard Alexis Larrañaga Rubio}

\affiliation{National University of Colombia}

\affiliation{Astronomical National Observatory (OAN)}

\begin{abstract}
As is already known, a spacetime horizon acts like a boundary of a
thermal system an we can associate with it notions as temperature
and entropy. Following the work of M. Akbar, in this paper we will
show how it is possible to interpret the field equation of a charged
BTZ black hole near horizon as a thermodynamic identity $dE=TdS+P_{r}dA+\Phi dQ$,
where $\Phi$ is the electric potential and $Q$ is the electric charge
of BTZ black hole. These results indicate that the field equations
for the charged BTZ black hole possess intrinsic thermodynamic properties
near horizon.
\end{abstract}
\maketitle

\section{Introduction}

The event horizon of a black hole acts as the boundary of the spacetime
because it blocks any physical information to flow out from the black
hole to the rest of world. Bekenstein \cite{a1} showed that the black
holes has non-zero entropy (because they withhold information), while
Hawking \cite{a2} showed that black holes emit thermal radiation
with a temperature proportional to its surface gravity at the black
hole horizon \begin{equation}
T=\frac{\kappa}{2\pi}\end{equation}

and with an entropy proportional to its horizon area \cite{a3},

\begin{equation}
S=\frac{A}{4G}.\end{equation}
This quantities are connected through the identity \begin{equation}
dE=TdS,\end{equation}
that is called \emph{first law of black hole thermodynamics} \cite{a1,a2,a3}.
When the black hole has other properties as angular momentum $J$
and electric charge $Q$ (e.g in the Kerr-Newman solution), the first
law can be generalized to

\begin{equation}
dE=TdS+\Omega dJ+\Phi dQ,\end{equation}
 where $\Omega=\frac{\partial M}{\partial J}$ is the angular velocity
and $\Phi=\frac{\partial M}{\partial Q}$ is the electric potential. 

The first law indicates that it could be possible to obtain a thermodynamic
interpretation of the Einstein field equations near horizon \cite{a5},
since the geometric quantities of the spacetime are related with the
thermodynamic quantities. This fact was used by Jacobson \cite{jac}
to find Einstein equations using the first law and the proportionality
between entropy and horizon area. On the other side, Paranjape et.al.
\cite{PSP} made an interpretation of the field equations as a thermodynamic
law $TdS=dE+PdV$ near the horizon of a special class of spherically
symmetric black hole. Kothawala et.al. \cite{ksp} extended this kind
of interpretation to stationary axis-symmetric horizons using the
virtual displacement of the horizon. This approach was used very recently
by M. Akbar \cite{Akbar} to make a thermodynamical interpretation
of the field equations of (2+1) gravity near the horizon of the BTZ
black hole.

In this paper we will use the same method to obtain a thermodynamical
identity using the field equation and considering the virtual displacement
of the horizon of a charged BTZ black hole.

\section{The Charged BTZ Black Hole}

The (2+1)-dimensional BTZ (Banados-Teitelboim-Zanelli) black holes
have obtained a great importance in recent years because the provide
a simplified model for exploring some conceptual issues, not only
about black hole thermodynamics \cite{btz,btz1} but also about quantum
gravity and string theory. 

The charged BTZ black hole is a solution of the (2+1)-dimensional
gravity theory with a negative cosmological constant $\Lambda=-1/\ell^{2}$
. The metric is given by\cite{chargedBTZ} \begin{equation}
ds^{2}=-f(r)dt^{2}+\frac{dr^{2}}{f(r)}+r^{2}d\varphi^{2},\label{eq:metric}\end{equation}
 where \begin{equation}
f(r)=-M+\frac{r^{2}}{\ell^{2}}+\frac{Q^{2}}{2}\ln\left[\frac{r}{\ell}\right]\end{equation}
 is known as the lapse function and $M$ and $Q$ are the mass and
electric charge of the BTZ black hole, respectively. The electric
potential of this black hole is 

\begin{equation}
\Phi=\frac{\partial M}{\partial Q}=-Q\ln\left[\frac{r}{\ell}\right].\label{eq:potential}\end{equation}
As it can be seen, the lapse function vanishes at the radii $r=r_{\pm}$,
where $r_{+}$ gives the position of the event horizon. The Bekenstein-Hawking
entropy of the BTZ black holes is twice the perimeter of theis event
horizon \cite{a17} \begin{equation}
S=4\pi r_{+},\end{equation}
 while the Hawking temperature is given, as usual, by \begin{equation}
T=\frac{1}{4\pi}\left|\frac{df(r)}{dr}\right|_{r=r_{+}}.\end{equation}
 These quantities obey the first law of thermodynamics, and we will
show, following the work of M. Akbar\cite{Akbar}, that the field
equations have the thermodynamic interpretation of a first law near
the horizon. In order to obtain this interpretation, we will assume
that the function $f(r)$ has a zero at $r=r_{+}$ and $f^{\prime}(r_{+})\neq0$
but has a finite value at $r=r_{+}$. These conditions assure that
we have a space-time horizon at $r=r_{+}$ and we can associate a
non-zero surface gravity $\kappa=\frac{1}{2}f'(r_{+})$ and a temperature
$T=\kappa/2\pi$ , while the associated entropy will be proportional
to the horizon area.

\section{Field Equations as a Thermodynamic Identity}

In $\left(2+1\right)$-dimensional gravity, the components of the
Einstein tensor $G_{\mu\nu}=R_{\mu\nu}-\frac{1}{2}Rg_{\mu\nu}$ for
the metric (\ref{eq:metric}) are simply \begin{equation}
G_{t}^{t}=G_{r}^{r}=\frac{f'(r)}{2r},\label{4}\end{equation}
 and \begin{equation}
G_{\varphi}^{\varphi}=\frac{f''(r)}{2},\label{6}\end{equation}
 where prime stands for the derivative with respect to $r$. It is
clear that $G_{0}^{0}$ and $G_{1}^{1}$ are equal and in particular,
at the horizon $r=r_{+}$, we have \begin{equation}
G_{0}^{0}|_{r=r_{+}}=G_{1}^{1}|_{r=r_{+}}=\frac{f'(r_{+})}{2r_{+}}.\label{7}\end{equation}
 The (2+1)-dimensional field equations are \begin{equation}
G_{\mu\nu}+\Lambda g_{\mu\nu}=-\pi T_{\mu\nu},\end{equation}
 where the units are such that $G=\frac{1}{8}$ and $c=1$. Here $T_{\mu\nu}$
is the stress-energy tensor, and it is such that $T_{r}^{r}=P_{r}$,
with $P_{r}$ the radial pressure. The $\left(r,r\right)$ component
of the field equations for this metric, when evaluated at $r=r_{+}$,
is \begin{equation}
\frac{1}{2a}f'(r_{+})-\frac{1}{\ell^{2}}=-\pi T_{r}^{r}.=-\pi P_{r}\end{equation}
 Now, we consider a virtual displacement $dr_{+}$ of the horizon
and we multiply both sides of this equation by it, \begin{equation}
\frac{f'(r_{+})}{4\pi}d(4\pi r_{+})-\frac{d(r_{+}^{2})}{\ell^{2}}=-P_{r}d(\pi r_{+}^{2}).\label{eq:aux1}\end{equation}
 The term $\frac{f'(a)}{4\pi}$ on the LHS is the associated temperature
$T$ while the companying quantity inside parentheses is the entropy
$S$ associated with the horizon. Using the condition $f\left(r_{+}\right)=0$
we have

\begin{equation}
-M+\frac{r_{+}^{2}}{\ell^{2}}+\frac{Q^{2}}{2}\ln\left[\frac{r_{+}}{\ell}\right]=0,\end{equation}

and then we obtain

\begin{equation}
-dM+d\left(\frac{r_{+}^{2}}{\ell^{2}}\right)+\frac{Q^{2}}{2r_{+}}dr_{+}=0,\end{equation}

\begin{equation}
d\left(\frac{r_{+}^{2}}{\ell^{2}}\right)=dM-\frac{Q^{2}}{2r_{+}}dr_{+},\end{equation}

that corresponds to the second term in the LHS of equation (\ref{eq:aux1}).
Thus, we have the equation

\begin{equation}
dM=TdS+P_{r}dA+\frac{Q^{2}}{2r_{+}}dr_{+},\end{equation}
 Here $dA$ is the change in horizon area and then, the term $P_{r}dA$
corresponds to work done against the pressure. Using the electric
potential given by (\ref{eq:potential}) we can write

\begin{equation}
\frac{Q}{r_{+}}dr_{+}=-dQ\ln\left[\frac{r_{+}}{\ell}\right],\end{equation}

so the field equation is

\begin{eqnarray}
dM & = & TdS+P_{r}dA-\frac{Q}{2}\ln\left[\frac{r_{+}}{\ell}\right]dQ\\
dM & = & TdS+P_{r}dA+\frac{\Phi}{2}dQ.\label{eq:aux3}\end{eqnarray}

As we can see, this equation resembles almost perfectly the first
law of thermodynamics but there is one more step left in order to
cancel the $\frac{1}{2}$ factor in the last term. Following the appreciation
of Martinez et.al.\cite{chargedBTZ} we can add on both sides f this
equation the term $-\frac{1}{2}Q\ln\left[\frac{r_{+}}{\ell}\right]dQ=\frac{1}{2}\Phi dQ$,
that is jut the electrostatic energy outside a sphere of radius $r_{+}$
(in our case this sphere is the horizon of the black hole). Thus,
equation (\ref{eq:aux3}) is

\begin{equation}
dE=TdS+P_{r}dA+\Phi dQ,\label{eq:firstlaw}\end{equation}

where $E$ is given by

\begin{equation}
dE=dM--\frac{1}{2}Q\ln\left[\frac{r_{+}}{\ell}\right]dQ,\end{equation}

and can be thought as the total energy within the radius $r_{+}$.
Hence, the field equation near horizon of the charged BTZ black hole
can be expressed as the thermodynamic identity\ref{eq:firstlaw} under
the virtual displacement of the horizon.

\section{Conclusion}

We have shown that the field equations for the charged BTZ black hole
have a thermodynamic interpretation near the horizon. As it has been
shown, we obtained the first law $dE=TdS+P_{r}dA+\Phi dQ$, where
$P_{r}$ is the radial pressure of the source, $A$ is the area enclosed
by the horizon, $\Phi$ is the electric potential, $Q$ is the electric
charge and $E$ is the total energy inside the horizon of the charged
BTZ black hole, that is the mass of the black hole minus a term that
corresponds to the electrostatic energy outside the horizon. The charged
rotating BTZ black hole case can be considered for further investigation.

\section*{Acknowledgments}

I would like to thanks Dr. J. M. Tejeiro for his useful discussion
and comments.


\begin{thebibliography}{10}
\bibitem{Akbar}M. Akbar. \emph{Thermodynamic Interpretation of Field
Equations at Horizon of BTZ Black Hole.} Chin. Phys. Lett. \textbf{24,
5.} (2007)1158.

\bibitem{chargedBTZ}C. Martinez, C. Teitelboim and J. Zanelli. \emph{Charged
Rotating Black Hole in Three Spacetime Dimensions.} Phys. Rev. \textbf{D61}
(2000) 104013  \emph{}\textbf{hep-th/9912259}

\bibitem{a2} S. W. Hawking, Commun. Math. Phys. \textbf{43}, 199
(1975). 

\bibitem{jac} T. Jacobson, Phys. Rev. Lett. \textbf{75}(1995)1260. 

\bibitem{btz} S. Carlip, Class. Quant. Grav. \textbf{12}(1995)2853. 

\bibitem{btz1} A. Ashtekar,Adv. Theor. Math. Phys. \textbf{6}(2002)507. 

\bibitem{a1} J. D. Bekenstein, Phys. Rev. D\textbf{7}, 2333 (1973);
Phys. Rev. D\textbf{9}, 3293 (1974). 

\bibitem{a3} J. M. Bardeen, B. Carter and S. W. Hawking, Commun.
Math. Phys. \textbf{31}, 161 (1973). 

\bibitem{PSP}A.~Paranjape, S.~Sarkar and T.~Padmanabhan, Phys.
Rev. D\textbf{74}, 104015(2006) \textbf{hep-th/0607240}.

\bibitem{a5} T. Padmanabhan, Mod. Phys. Letts. A\textbf{17}, 923
(2002) \textbf{gr-qc/0202078}. Phys. Rept. \textbf{406}, 49 (2005)
\textbf{gr-qc/0311036}. 

\bibitem{a17} M. Banados, C. Teitelboim and J. Zanelli, Phys. Rev.
Lett. \textbf{69}, 1849(1992);M. Banados, M. Henneaux, C. Teitelboim
and J. Zanelli, Phys. Rev. D \textbf{48}, 1506(1993). 

\bibitem{ksp} D. kothawala, S. Sarkar, and T. Padmanabhan, \textbf{gr-qc/0701002}. 
\end{thebibliography}
\end{document}